\documentclass[12pt]{article}
\usepackage{xcolor}
\usepackage{amsmath}
\usepackage{graphicx}
\usepackage{times}
\usepackage{geometry}
\usepackage[utf8]{inputenc}
\usepackage{enumitem,amssymb}
\usepackage{ragged2e}
\usepackage{cite}
\newlist{thematic}{itemize}{8}
\setlist[thematic]{label=$\square$}
\usepackage{pifont}

\usepackage{natbib}
\usepackage{lscape}
\usepackage{authblk}
\usepackage{hyperref}
\usepackage{indentfirst}

\newcommand{\solphys}{Solar~Phys.}
\newcommand{\apj}{Astrophys.~J.}

\begin{document}
\raggedright
\huge
Helio2024 Science White Paper \linebreak

{Solar and Heliospheric Magnetism in 5D.}\linebreak
\normalsize
  
\textbf{Principal Author:} Alexei A. Pevtsov (National Solar Observatory) 
 
\textbf{Co-authors:}
 T. Woods (University of Colorado, LASP),
 V. Martinez-Pillet (National Solar Observatory), 
 D. Hassler (SwRI), 
 T. Berger (University of Cororado, SWx TREC), 
 S. Gosain (National Solar Observatory), 
 T. Hoeksema (Stanford University), 
 A. R. Jones (University of Colorado, LASP),  
 R. Kohnert (University of Colorado, LASP), 
 T. Y. Chen (Columbia University), 
 L. Upton (SouthWest Reseatch Institute), 
 A. Pulkkinen (NASA, Goddard Space Flight Center) 
\linebreak
\linebreak
\vskip 2cm
\textbf{SYNOPSIS:}
This White Paper argues for the urgent need for the multi-vantage/multi-point observations of the Sun and the heliosphere in the framework of six (6) key science objectives. We further emphasize the critical importance of 5D-``space'': three spatial, one temporal and the magnetic field components. The importance of such observations cannot be overstated both for scientific research and the operational space weather forecast.

\newpage

\section{Introduction}
Magnetic field in the solar atmosphere and heliosphere is a global, ever-changing, multi-scale system. Active regions that emerge, evolve, and decay in one “place” on the solar surface may cause small or big changes in other remote areas and in the extreme cases, over the whole solar corona. Small-scale instabilities could cause localized eruptions, or they may cascade to much larger scales, both across the “surface” and with height.  Once the magnetized structures start erupting from the solar atmosphere, their magnetic systems may go through a complex reconnection process with overlaying large-scale fields including those rooted in solar polar areas. After it erupts, the magnetic system continues evolving as it travels towards Earth, Mars and beyond. In addition to spatial scales, magnetic fields may evolve on different time scales from rapid eruption processes to relatively slow evolutionary changes. To properly capture and study these changes in different spatial and temporal scales requires taking observations from multiple vantage points at sufficiently high time cadence, which we refer to as 5D concept (3D for three spatial directions, time, and the vector magnetic field). 

\section{Science Objectives}
The following six key inter-related science objectives are important to address to advance the understanding of solar and heliospheric magnetism in 5D. 

\subsection{Understand the global interconnected magnetic system in the solar corona}

Active region emergence may cause magnetic field restructuring in remote locations 
on the Sun \citep{Zhang.Low2001,Zhang.Low2002,Longcope.etal2005}. These topological changes may result in the formation and/or disappearance of coronal holes \citep{Karachik.etal2010}, or remote and even global triggering of flares and filament/CME eruptions \citep{Balasubramaniam.etal2011,Fu.Welsch2016,Schrijver.Title2011}. Understanding of such global connectivity and its evolution is virtually impossible without multi-vantage observations of solar surface. Moreover, this global magnetic system extends with height and distance from the sun into upper corona and heliosphere. Having a multi-vantage observations of magnetic fields on the sun combined with state-of-the-art modeling would allow reconstructing a more realistic topological structure of solar corona, but to understand the 3D structure of heliosphere between sun and 1AU also requires taking \emph{in situ} magnetic measurements from multi-vantage points. Hence, our second key objective:

\subsection{Understand the evolving structures of ICMEs and solar wind streamers}

Our current understanding of the temporal and spatial structure of interplanetary phenomena (ICMEs\footnote{ICME --- Interplanetary Coronal Mass Ejection} and fast solar wind streamers) is based on \emph{in situ} single point measurements of magnetic field and solar wind.  A spacecraft is in some point in 3D space, and it takes measurements along a line crossing a fast-moving heliospheric “disturbance” (e.g., ICME, solar wind stream).  The orientation of this line of measurements in respect to the overall structure of disturbance is unknown, and thus, to infer the 3D structure of ICMEs requires making additional assumptions about the topology of its magnetic field (e.g., spheromak or twisted flux tube like structure, and/or  using statistical tendencies of past events). Multi-point \emph{in situ} measurements would allow to create true topological model of these interplanetary structures and significantly improve their modeling and the prediction of space weather events.
Our first key objective refers to a global magnetic connectivity in solar corona. However, this connectivity is highly dynamic. Taking simple “snap-shots” is likely to miss the role of dynamic changes. Tracking key evolutionary changes requires taking observations from a sufficient number of vantage points along ecliptic plane. At a minimum, it could be four viewing angles, for example, corresponding to Lagrangian L1, L4, L5, L3 points, or a similar number of drifting spacecraft on STEREO-like orbits. In addition to tracking the evolutionary changes due to global restructuring the corona, these observations would also allow studying the evolution of major “building blocks” of solar activity (e.g., sunspots and active regions, coronal holes, chromospheric filaments, etc). Our third key objective is one motivation for such studies.

\subsection{Understand the full life-cycle of active region emergence, growth, and decay}

Sunspots and active regions (ARs) are hallmark of solar magnetic activity. However, observations from a single vantage point (Earth) are limited to about 10 days for accurately measuring magnetic field of 
a single AR and is thus insufficient for observing the emergence of most ARs and for studying the complete life-cycle of individual ARs. Without observing the complete life-cycle of ARs we cannot fully quantify  their fundamental properties such as the dispersion rate of active region magnetic flux, the rates of and balance between flux eruption and decay in active regions, true active region magnetic topology including inter-connectivity with magnetic fields outside the active region and between open and closed configurations.
Using vector magnetic field measurements could help in understanding the 3D topology magnetic fields even from a single vantage point (e.g., L1 or ground-based observatory). However, the interpretation of vector fields has one important limitation, which multi-vantage points could help to address in our fourth objective.

\subsection{Understand the true 3D orientation of magnetic fields and their role in eruptive events}

Understanding of true topology of magnetic fields in the solar atmosphere and derivation of high-level topological invariants (e.g., helicity) critical for understanding the solar dynamo and magnetic reconnection requires observations of vector magnetic fields. Knowledge of true orientation of magnetic fields in magnetic systems interacting with each other (e.g., due to their emergence or growth) is critical for understanding how the magnetic fields reconnect and erupt. However, vector field observations have one intrinsic limitation – 180-degree ambiguity in orientation of the horizontal magnetic field. Multi-vantage observations would enable resolution of this 180-degree ambiguity, and the determination of proper 3D orientation of magnetic fields in solar corona.
Modern state-of-the-art modeling of solar wind and ICMEs strongly suggest that the knowledge of magnetic fields in solar polar areas is critical for understanding the global magnetic connectivity on the sun. Ultimately, above-the-pole views are required to accurately measure
the 
polar magnetic fields. However, multi-vantage observations from the ecliptic plane can also be advantageous for long-term tracking of solar polar fields as is important for our key objective 5: 

\subsection{Explore the roles and processes that solar polar magnetic fields have on the solar global fields and solar dynamo}

Uncertainty in the magnetic flux (polarity and amplitude) in solar polar regions is the largest contributor to the errors and uncertainties in model predictions of global fields, that consequently degrade the forecast accuracy for 
solar wind structure and CME arrival times. From a single vantage point in the ecliptic plane (e.g., Earth), each solar pole is only visible for about 6 months out of the year, and then only with severe distortion due to the grazing angles of our line-of-sight to the poles.  Lack of continuous observations of solar polar magnetic fields also contributes to so called “missing” magnetic flux (open flux problem), when the total magnetic flux derived from \emph{in situ} measurements is significantly smaller than the magnetic flux observed on solar surface. The solar polar fields are also critical for advancing the understanding of the solar dynamo processes that move magnetic fields on and within the Sun over many years and is thought to have the largest influence on the 11-year solar activity cycle (22-year magnetic cycle). The deep and shallow dynamos are poorly constrained today because of the limited knowledge of the solar polar fields.
Observations from multiple spacecraft located at different azimuths along the ecliptic plane could provide views of polar regions different from the L1/Earth view, potentially improving our understanding of the solar poles. However, the only way to truly measure, monitor, and incorporate accurate polar fields into solar wind models is to have a system of out-of-the-ecliptic spacecraft which continually provide “look down” observations of the solar poles. Solar Orbiter may achieve high-inclination views of the solar poles in the 2030 time frame, but the data rate and instrumentation limitations of that mission ensure that it will be a limited but tantalizing glimpse of what we would find with a true polar orbiting mission. Such a mission imposes much more complex launch and operation requirements compared to ecliptic-plane missions. But as shown by the out-of-the-ecliptic observations of the polar regions of the giant planets Jupiter and Saturn from the Juno and Cassini missions, respectively, major discoveries are very likely to result when we look straight down on the polar regions of our central star.  A combination of multi-vantage ecliptic plane observations and occasional over-the-pole observations would allow studying the structure of polar areas and its long-term evolution.
 Although this white paper concentrates exclusively on the magnetic fields, helioseismology from multiple vantage points also improves the sensitivity to flows and fields. Our final key objective refers to the improvements to the space weather research and operational forecast:

\subsection{Improve prediction of solar wind structure and CME impacts throughout the Heliosphere} 

Multi-point observations of the Sun and throughout the interplanetary space would revolutionize operational space weather forecasting. It would also allow extending the forecast to other locations throughout the solar system (e.g., Mars, Earth-Mars transfer orbits, etc). The above advantages of multi-vantage observations of solar surface and the heliosphere needs to be quantified. This could be done using the modern modeling approaches and model-to-model comparison. As one example, we refer to \citet{Pevtsov.etal2020}, who analyzed the improvements in solar wind prediction as compared with “ground truth” when using four-vantage point observations (L1+L4+L5+L3) as compared with one  (L1), two (L1+L5), and even three (L1+L5+L4) vantage points. They also show that a single out of ecliptic observation (e.g., Solar Orbiter) does not result in a sufficient improvement in all places along the ecliptic plane. Such modeling is a critical element for planning future multi-vantage point missions.

\section{Recommendation}

Multi-vantage observations of Sun and the heliosphere have strong potential in revolutionizing our understanding of solar active phenomena and their effect on planetary magnetospheres (space weather). It is desirable to gradually build-up a system of ground-based and space-based instruments to provide comprehensive datasets for developing a complete understanding of  magnetic environment around our Sun in space and time.

\end{document}